\documentclass[apj]{emulateapj}
\usepackage{apjfonts}

\usepackage{graphicx}
\received{January 7, 2010}
\accepted{April 15, 2010}

\newcommand\teff{\ensuremath{T_{\rm eff}}}

\slugcomment{Accepted for publication in the Astrophysical Journal} 

\shorttitle{Temperature Change and Oblique Pulsation in GD~358}

\shortauthors{Montgomery et al.}

\begin{document}
\title{Evidence for Temperature Change and Oblique Pulsation from
  Light Curve Fits of the Pulsating White Dwarf GD~358}

\author{
  M.\ H.\ Montgomery\altaffilmark{1,3}, 
  J. L. Provencal\altaffilmark{2,3}, 
  A. Kanaan\altaffilmark{4},  
  Anjum S. Mukadam\altaffilmark{5},   
  S. E. Thompson\altaffilmark{2,3}, 
  J. Dalessio\altaffilmark{2},
  H. L. Shipman\altaffilmark{2}, 
  D. E. Winget\altaffilmark{1}, 
  S. O. Kepler\altaffilmark{6},
  and
  D. Koester\altaffilmark{7}
}

\altaffiltext{1}{Department of Astronomy and McDonald Observatory, 
  University of Texas at Austin, Austin, TX, USA; mikemon@astro.as.utexas.edu}
\altaffiltext{2}{Department of Physics and Astronomy, University of 
  Delaware, Newark, DE, USA}
\altaffiltext{3}{Delaware Asteroseismic Research Center, Mt.\ Cuba
  Observatory, Greenville, DE, USA} 
\altaffiltext{4}{Departamento de F\'{\i}sica Universidade Federal de Santa Catarina, C.P. 476, 88040-900, Florian{\'o}polis, SC, Brazil}
\altaffiltext{5}{Department of Astronomy, University of Washington, Seattle, WA, USA}
\altaffiltext{6}{Instituto de F\'{\i}isica UFRGS, C.P. 10501, 91501-970 Porto Alegre,RS, Brazil}
\altaffiltext{7}{Institut f\"ur Theoretische Physik und Astrophysik, Universit\"at Kiel, 24098 Kiel, Germany}

\begin{abstract}
  
  Convective driving, the mechanism originally proposed by
  \citet{Brickhill91a,Brickhill83} for pulsating white dwarf stars,
  has gained general acceptance as the generic linear instability
  mechanism in DAV and DBV white dwarfs.  This physical mechanism
  naturally leads to a nonlinear formulation, reproducing the observed
  light curves of many pulsating white dwarfs.  This numerical model
  can also provide information on the average depth of a star's
  convection zone and the inclination angle of its pulsation axis. In
  this paper, we give two sets of results of nonlinear light curve
  fits to data on the DBV GD~358. Our first fit is based on data
  gathered in 2006 by the Whole Earth Telescope (WET); this data set
  was multiperiodic, containing at least 12 individual modes. Our
  second fit utilizes data obtained in 1996, when GD~358 underwent a
  dramatic change in excited frequencies accompanied by a rapid
  increase in fractional amplitude; during this event it was
  essentially monoperiodic. We argue that GD~358's convection zone was
  much thinner in 1996 than in 2006, and we interpret this as a result
  of a short-lived increase in its surface temperature.  In addition,
  we find strong evidence of oblique pulsation using two sets of
  evenly split triplets in the 2006 data. This marks the first time
  that oblique pulsation has been identified in a variable white dwarf
  star.

\end{abstract}

\section{Astrophysical Context}

White dwarf stars offer several advantages for astrophysical study.
First, they are the evolutionary endpoint of about 97\% of all stars
and are therefore representative of a large fraction of the stellar
population. Second, the source of their pressure support is electron
degeneracy \citep{Chandrasekhar39} so their bulk mechanical structure
is well understood. Third, nuclear reactions, if any, contribute a
negligible amount to their energy, so their evolution is dominated by
simple cooling \citep{Mestel52}. Finally, they are observed to pulsate
in specific temperature ranges. The pulsators are believed to be
typical in every other way, so what we learn
\emph{asteroseismologically} about them should apply to all white
dwarf stars \citep[for recent reviews, see][]{Winget08,Fontaine08}.

In addition to learning about the stars themselves, the relative
simplicity of white dwarfs makes them ideal laboratories for testing
and constraining poorly-understood physical processes. One such
process, convection, is an important energy transfer process in most
stars, yet it remains one of the largest sources of uncertainty in
stellar modeling. For instance, main-sequence stars at least 20\% more
massive than the Sun have convective cores, and the amount of
convective overshoot and mixing is the primary factor that determines
their main sequence lifetimes \citep[see, e.g.,][]{DiMauro03}.  In
addition, red giants and AGB stars have large convective envelopes,
and the details of convection play a role in the evolution of their
surface abundances and in their overall evolution \citep{Bertelli09}.

We have developed a method which uses the pulsations of white dwarf
stars to measure fundamental parameters of their convection zones.
The physical idea is that the pulsations cause local surface
temperature (``\teff'') variations that lead to local variations in
the depth of the convection zone. As the convection zone waxes and
wanes it both absorbs and releases energy, modulating the local energy
flux \citep{Brickhill91a,Goldreich99a}. Due to the extreme temperature
sensitivity of convection, finite amplitude pulsations can lead to
highly nonlinear light curves \citep{Brickhill92a,Wu01,Ising01}.  In
\citet{Montgomery05a} we showed how a simple numerical model could be
used to obtain not only good light curve fits but also information on
the average depth of a star's convection zone and the inclination
angle of its pulsation axis.

\section{Nonlinear Light Curve Fits}

\citet{Montgomery05a} demonstrated that by considering the nonlinear
response of the convection zone \citep{Brickhill92a,Wu01} one could
obtain excellent fits to the light curves of two (nearly) single mode
white dwarf pulsators. We have since extended this technique to
multi-periodic stars \citep{Montgomery07a} and have taken into account
the nonlinear relationship between the bolometric and observed flux
variations \citep{Montgomery08a}. In this section we describe these
effects and show how they have been added to our nonlinear light curve
fitting technique.

Our approach is well-summarized in \citet{Montgomery05a} and
\citet{Montgomery08a}. Briefly, we assume that the convective turnover
time of fluid elements in the convection zone is short ($\la 1$~s)
compared to the periods of pulsation ($\ga 100$~s), so that the
convection zone responds almost instantaneously to the pulsations and
is always in hydrostatic equilibrium. In addition, we assume that the
perturbation of the flux at the base of the convection zone is
sinusoidal. Next, we assume that the luminosity changes are due only
to temperature changes and \emph{not} to geometric effects;
\citet{Robinson82} found that the fractional radius change $\Delta R/R
\sim 10^{-5}$--$10^{-4}$ for the DAVs, with temperature variations
(and the associated changes in limb darkening) accounting for well
over 99\% of the luminosity variation, and this result was confirmed
by \citet{Watson88} for both the DAV and DBV stars. Finally, the
surface convection zones of pulsating white dwarfs are quite thin
geometrically, of order $10^{-5}$ of the radius of the star; this
allows us to use the plane parallel approximation and neglect
horizontal energy transport.  With these assumptions we are able to
derive a relationship between the local flux at a given ($\theta$,
$\phi$) entering the convection zone at its base, $F_{\rm base}$, and
that leaving it at the photosphere, $F_{\rm phot}$:
\begin{equation}
  F_{\rm phot} = F_{\rm base} + \tau_C \frac{dF_{\rm phot}}{dt},
\end{equation}
where the new timescale $\tau_C\equiv\tau_C(F_{\rm phot})$ describes
the changing heat capacity of the convection zone as a function of the
local photospheric flux. We parameterize it as
\begin{equation}
  \tau_C = \tau_0 \left(\frac{T_{\rm eff}}{T_{\rm eff,0}}\right)^{-N},
\end{equation}
where $\tau_0$ is the equilibrium value of $\tau_C$, $T_{\rm eff}$ is
the instantaneous effective temperature and $T_{\rm eff,0}$ is its
equilibrium value, and $N$ is a parameter describing the sensitivity
of $\tau_C$ to changes in $T_{\rm eff}$. From standard mixing length
theory of convection we expect that $N \sim 90$ for DAVs and $N \sim
23$ for DBVs. It is this extreme temperature sensitivity which is
responsible for the large nonlinearities seen in white dwarf
pulsations.  For reference, this timescale is closely related to the
standard thermal timescale ($\tau_{\rm th}$) at the base of the
convection zone: for the DAVs $\tau_C \approx 4 \,\tau_{\rm th}$
\citep{Goldreich99a} and for the DBVs $\tau_C \approx 0.6\,\tau_{\rm
  th}$. With the further assumption that the angular dependence of
$F_{\rm base}$ is given by a spherical harmonic $Y_{\ell m}$, we can
calculate the bolometric flux changes at the surface of the model, and
average them appropriately over the visible disk of the model.

\begin{figure}[t]
  \centering{\includegraphics[width=\columnwidth]{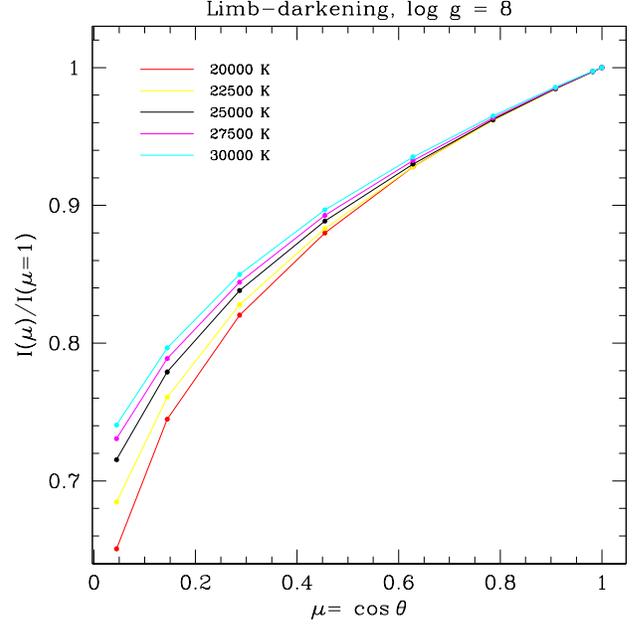}}
  \caption{An example of the limb-darkening in DB model atmospheres
    with $\log g = 8.0$, for the indicated range of temperatures.}
  \label{limb}
\end{figure}

\subsection{Improvements in the Modeling}

Since 2005 we have made important technical improvements to the light
curve fitting code.  First, we extended the code to include the more
common multiperiodic case, where many modes with different $\ell$ and
$m$ values are simultaneously present. Thus, the flux at the base of
the convection zone is now given by a sum over the modes:
\begin{equation}
\frac{\delta F_{\rm base}}{F_{\rm base}} =
      {\rm Re} \left\{ \sum_{j=1}^M A_j e^{i (\omega_j t+\phi_j)}
        Y_{\ell_j m_j}(\theta,\phi)
        \right\}.
\end{equation}
In this formula, $A_j$, $\omega_j$, $\phi_j$, $\ell_j$, and $m_j$ are
the amplitude, angular frequency, phase, $\ell$, and $m$ values of the
$j$-th mode, and the total number of modes is $M$.

Second, we adapted the code to simultaneously fit an arbitrary number
of observations (``runs'').  This is a necessary step for applying
this technique to multiple runs obtained from Whole Earth Telescope
(WET) campaigns as well as successive nights of single-site
observations. Since the code allocates and deallocates memory as
needed it typically uses only 8 MB of RAM, independent of the number
of runs included in the fit.

Third, we replaced our simple analytical prescription for limb
darkening with tabulated values based on a grid of our model
atmospheres \citep[for a description of the models, see][]{Koester10}.
This grid ranges in \teff\ from 20,000~K to 30,000~K in steps of 500~K
and in $\log g$ from 7.5 to 8.5 in steps of 0.25. From this grid we
instantaneously calculate the local flux as a function of \teff\ and
$\mu \equiv \cos \theta$; thus, variations in the limb darkening with
\teff\ are automatically included. In Figure~\ref{limb} we show
examples of the limb darkening for a $\log g = 8.0$ He atmosphere
white dwarf model as a function of $\mu$ for a range of \teff\ 
values.

Finally, we improved the way in which the local bolometric flux
variations are mapped into variations in a given wavelength band.
Previously we used a flux ``correction factor,'' $\alpha_X$, to
accomplish this. Denoting by $F_X$ the flux in the passband $X$, then,
for small fractional changes in the fluxes, $\alpha_X$ was defined by
\begin{equation}
  \frac{\delta F_X}{F_X} = \alpha_X \frac{\delta F_{\rm phot}}{F_{\rm phot}},
  \label{alphadef}
\end{equation}
where $F_{\rm phot}$ is the local bolometric flux at the photosphere
and $\delta F$ is the variation in the respective fluxes due to the
pulsations. Clearly, $\alpha_X$ depends on the wavelength coverage of
the passband as well as the wavelength distribution of the flux from
the source. In previous analyses we estimated that $\alpha \sim 0.42$
for DBVs and $\alpha \sim 0.66$ for DAVs. The value for the DAVs is
not that different from what one obtains from a proper calculation
assuming a passband centered on 5000~\AA. For the DBVs, however, the
more detailed calculations yield a value of $\alpha \sim\, $0.25--0.35
which \emph{is} significantly different from the earlier estimates.

Of course, $\alpha$ is not strictly a constant but rather is a
function of \teff\ and therefore $F_{\rm phot}$.  For larger
amplitudes, departures from linearity between the fluxes become more
important, and to do the problem properly we need $F_X$ as a function
of $F_{\rm phot}$, i.e., $F_X \equiv F_X(F_{\rm phot})$.

To calculate this flux conversion we use the same grid of DB model
atmospheres described earlier.  In each model atmosphere, the flux is
tabulated as a function of wavelength $\lambda$ and viewing angle
$\mu$, so limb darkening is automatically included. Using a given
passband, we integrate it against the model atmosphere flux to obtain
$F_X$ as a function of \teff, $\mu$, and $\log g$. Since $\log g$ is
essentially constant during the pulsations, we only have to
interpolate once between sets of models for the given value of $\log
g$; thus, we have $F_X\equiv F_X(\teff,\mu)$. Finally, since $F_{\rm
  phot} \propto \teff^4$, we can express this as $F_X\equiv F_X(F_{\rm
  phot},\mu)$.

In practice, we are only interested in relative flux changes, so we
calculate everything relative to a reference flux, which we take to be
the equilibrium flux of the star. If we denote by a subscript ``0''
the equilibrium values of the fluxes, then the function $f$ we want is
defined by
\begin{equation}
  \frac{F_X}{F_{X,0}} = 
     f\left(\frac{F_{\rm phot}}{F_{\rm phot,0}},\mu\right).
     \label{ffunc}
\end{equation}

\begin{figure}[t]
  \centering{
    \includegraphics[width=1.0\columnwidth,angle=0]{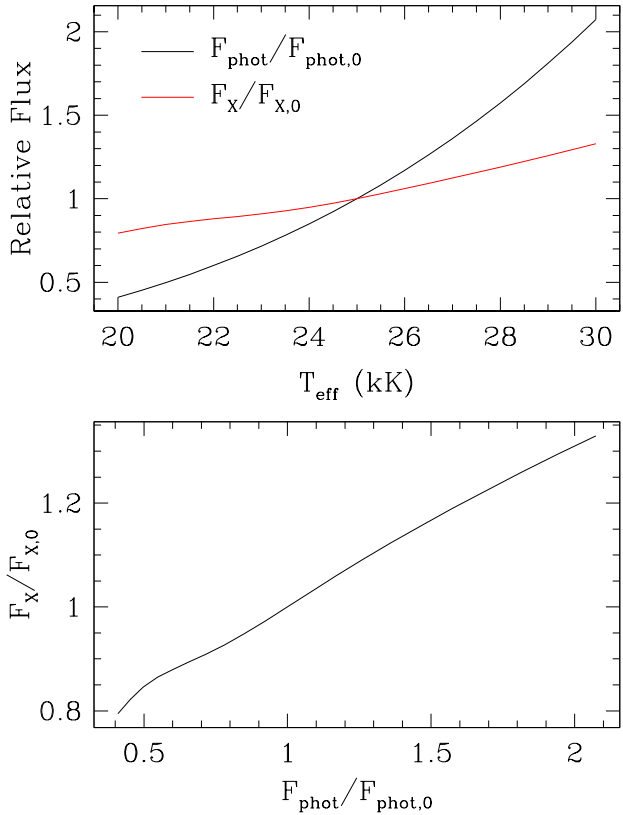}
    }
    \caption{Top panel: the bolometric flux ($F_{\rm phot}$) and the 
      flux in the given passband ($F_X$) as a function of \teff\ for
      $\mu=1$.  These calculations are for a $\log g = 8.0$ DB model,
      and both fluxes have been normalized to one at $\teff =
      25,000$~K. Lower panel: the passband flux as a function of the
      bolometric flux, again normalized at $\teff = 25,000$~K. This is
      the function $f$ given by equation~\ref{ffunc}.  }
    \label{fplot}
\end{figure}

In the top panel of Figure~\ref{fplot} we show the fluxes
$F_X/F_{X,0}$ and $F_{\rm phot}/F_{\rm phot,0}$ as a function of
\teff, both normalized to one at $\teff = 25,000$~K; the filter $X$ is
assumed to have an effective wavelength of $\sim 5200$~\AA.  The
response of this filter was obtained from convolving the wavelength
response of the ALFOSC-FASU CCD, the NOT primary and secondary mirror
reflectivities, an S8612 filter, and atmospheric absorption. In the
lower panel we show the function $f$ defined in equation~\ref{ffunc}
as derived from these calculations.  We see that while this is not a
perfectly linear relationship, the deviations from linearity are not
dramatic. Thus, while we use the fully nonlinear relation in our
calculations, we expect this effect to make only a minor contribution
to the overall nonlinearities associated with the pulsations.

\begin{deluxetable}{lccc}
\tablecolumns{4}
\tablewidth{0pt}
\tablecaption{Observing Runs Used for Light Curve Fits}
\tablehead{
\colhead {Run Name} &
\colhead {Telescope} &
\colhead {Instrument} &
\colhead {Length (hrs)}
}
\startdata
chin20060527 & BAO 2.16m & PMT & 4.5 \\ 
chin20060528 & BAO 2.16m & PMT & 5.2 \\ 
chin20060531 & BAO 2.16m & PMT & 3.9 \\ 
hawa20060518 & 0.6m & Apogee & 2.2  \\ 
hawa20060519 & 0.6m & Apogee & 1.7  \\ 
hawa20060520 & 0.6m & Apogee & 3.5  \\ 
hawa20060521 & 0.6m & Apogee & 7.8  \\ 
hawa20060522 & 0.6m & Apogee & 6.6  \\ 
hawa20060523 & 0.6m & Apogee & 3.7  \\ 
hawa20060524 & 0.6m & Apogee & 5.7  \\ 
hawa20060525 & 0.6m & Apogee & 8.8  \\ 
hawa20060526 & 0.6m & Apogee & 9.1  \\ 
hawa20060527 & 0.6m & Apogee & 7.8  \\ 
hawa20060528 & 0.6m & Apogee & 9.1  \\ 
hawa20060530 & 0.6m & Apogee & 8.7  \\ 
kpno20060518 & KPNO 2.1m & Apogee & 7.0 \\ 
kpno20060519 & KPNO 2.1m & Apogee & 7.3 \\ 
kpno20060520 & KPNO 2.1m & Apogee & 7.6 \\ 
kpno20060521 & KPNO 2.1m & Apogee & 7.3 \\ 
kpno20060522 & KPNO 2.1m & Apogee & 1.0 \\ 
kpno20060523 & KPNO 2.1m & Apogee & 2.9 \\ 
mcdo20060523 & 2.1m & Argos & 7.4  \\ 
mcdo20060524 & 2.1m & Argos & 7.2  \\ 
mcdo20060525 & 2.1m & Argos & 6.4  \\ 
mcdo20060528b & 2.1m & Argos & 7.2  \\ 
mcdo20060529 & 2.1m & Argos & 8.2  \\ 
nord20060607 & 2.7m & ALFOSC & 7.1  \\ 
nord20060608 & 2.7m & ALFOSC & 8.0  \\ 
nord20060609 & 2.7m & ALFOSC & 7.9  
\enddata
\label{obslist}
\end{deluxetable}
\section{Light Curve Fits to the 2006 WET Run}
\label{lcfits}

Our recent work with light curve fitting has been limited to nearly
single mode pulsators: the DAVs G29-38\footnote{More precisely, while
  G29-38 is normally multiperiodic, the data set used by
  \citet{Montgomery05a} was taken at a time when its light curve was
  dominated by a single large mode.}  and GD154, and the DBV PG1351+489.
This is because (1) mono-periodic data can be folded at the pulsation
period, producing a high S/N ``light curve,'' and (2) the number of
possible mode identifications ($\ell$ and $m$ values) for a single
mode is small enough that all possibilities can be directly explored.

\begin{figure*}[t]
  \centering{\includegraphics[height=0.92\textwidth,angle=-90]{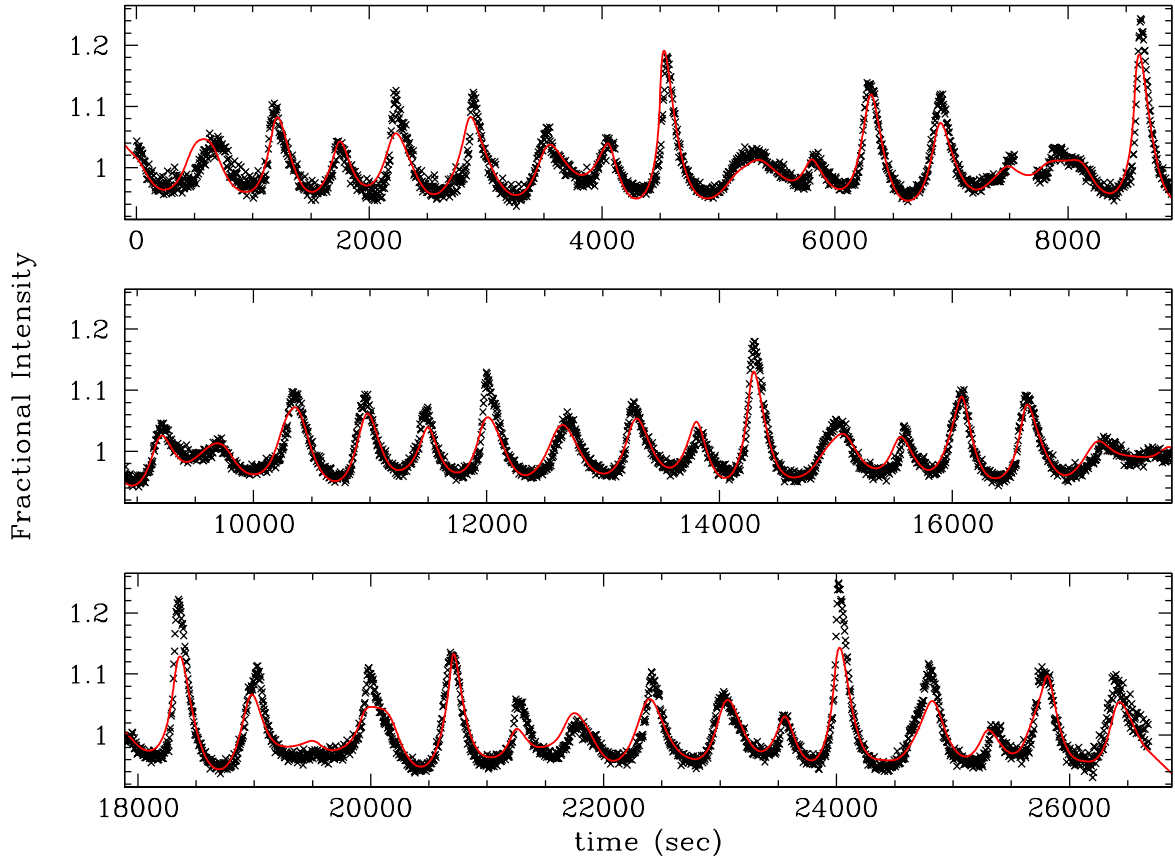}}
  \caption{A comparison of the simultaneous fit of Table~\ref{fitall}
    (solid line) to the light curve from run
    mcdo20060523 (crosses).  }
  \label{mcd23t}
\end{figure*}

\begin{figure*}[t]
  \centering{\includegraphics[height=0.92\textwidth,angle=-90]{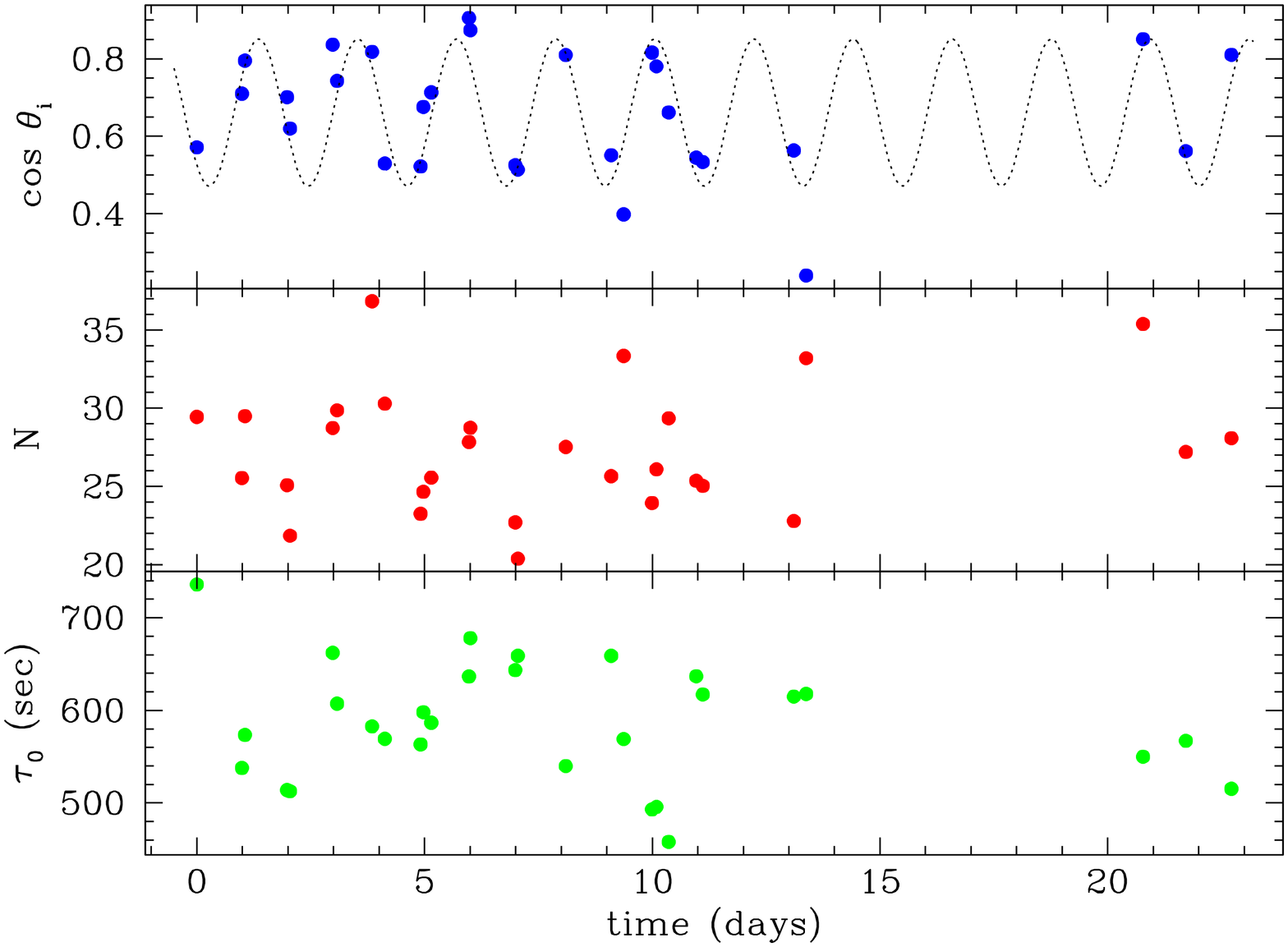}}
  \caption{The variation of $\theta_i$, $N$, and $\tau_0$ as a function
    of time during the 2006 WET run. While $\tau_0$ and $N$ show no
    obvious signatures, $\theta_i$ shows variations which may be
    sinusoidal in origin.  }
  \label{theta1}
\end{figure*}

GD~358 violates both of these conditions. First, due to the
nonlinear interaction of its large amplitude modes, the pulse shape
obtained by folding its light curve at a mode period is not the same
as the pulse shape which would be obtained in the absence of other
modes \citep{Montgomery07a}. Second, GD~358 has a large number of
observed modes (see Table~\ref{modelist}) and it is impractical to
search all possible combinations of $\ell$ and $m$ values which each
mode can take. For instance, if we assume GD~358 to have of order
$\sim 10$ modes, all of which have $\ell=1$, then all possible
permutations of $m$ values yield a number of $(2 \ell + 1)^{10} \sim
60,000$ different cases!  Since each fit takes of order an hour on a
single processor computer, this is completely impractical using a
standard desktop computing approach.

Fortunately, GD~358 is well studied, so we have a good idea what the
$\ell$ and $m$ values for the main pulsation modes are
\citep[e.g.,][]{Metcalfe00,Winget94}. Even so, our derived values of
$\tau_0$ depend only weakly on the assumed mode identifications, so it
is not necessary for the mode identification of each mode to be
exact. Furthermore, the extended time baseline and excellent
coverage of the 2006 WET run allows us to obtain very accurate
frequencies for these modes \citep{Provencal09}. With this as a basis,
our approach is to assume that the frequencies and mode
identifications are known and then to make nonlinear light curve fits
to a subset of runs in the WET campaign which have high S/N. This
implicitly assumes that the pulsations are coherent throughout the
time spanned by the runs; in section~\ref{obpm} we show that this is
not strictly true for some of the modes. High S/N data are desirable
since we are mainly interested in the nonlinear part of the light
curve, which itself is smaller than the linear part.

\begin{deluxetable}{ccrr}
\tablecolumns{4}
\tablewidth{13em}
\vspace*{2.0em}
\tablecaption{Independent Modes from the 2006 WET Run}
\tablehead{
\colhead {Period} & \colhead{$k$} &\colhead{$\ell$} &\colhead{$m$} \\
\colhead {(s)} &\colhead{} &\colhead{} &\colhead{}
}
\startdata
422.56 & 8 & 1 &  1 \\ 
423.90 & 8 & 1 & -1 \\ 
463.38 & 9 & 1 &  1 \\ 
464.21 & 9 & 1 &  0 \\ 
465.03 & 9 & 1 & -1 \\ 
571.74 & 12 & 1 &  1 \\ 
574.16 & 12 & 1 &  0 \\ 
575.93 & 12 & 1 & -1 \\ 
699.68 & 15 & 1 &  0 \\ 
810.29 & 18 & 1 &  0 \\ 
852.50 & 19 & 1 &  0 \\ 
962.38 & 22 & 1 &  0 
\enddata
\label{modelist}
\end{deluxetable}

Many high quality runs were taken during the May 2008 WET run, and we
list in Table~\ref{obslist} those used in our fits.  We included the
12 largest amplitude periodicities from the 2006 WET run which were
deemed to be independent frequencies and not linear combinations, and
these are given in Table~\ref{modelist}.  The $\ell$ and $m$
identifications are taken from previous analyses of this star
\citep{Winget94,Kepler03,Metcalfe03,Provencal09} and are based on
asymptotic theory as well as genetic algorithm fits. To calculate the
conversion from bolometric to observed passband fluxes, we must assume
values for \teff\ and $\log g$.  We chose the values of
\citet{Beauchamp99}, $T_{\rm eff} = 24900$~K and $\log g = 7.91$,
although we have also used those of \citet{Castanheira05} ($T_{\rm
  eff} = 24100$~K, $\log g = 7.91$) to assess the uncertainties this
introduces.

We began the fitting process with the highest S/N data taken with the
2.7m Nordic Telescope (NOT).  However, it became apparent that GD~358's
complex pulsation spectrum required an extended baseline of data to
constrain the phases of closely spaced frequencies.  Fortunately,
GD~358's brightness and large amplitude meant that a large number of
individual observing runs met the required S/N.

Our simultaneous fit to the runs in Table~\ref{obslist} yields the
following parameters: $ \tau_0 = 572.9 \pm 6.1 $ s, $ N = 23.5 \pm
0.1 $, $ \theta_i = 50.5 \pm 0.2^\circ$.  Figure~\ref{mcd23t} shows
the ability of the fit to reproduce the essential features of the
light curve (run mcdo20060523 plotted).  Additional results for this
fit, including the amplitude and phase for each mode, are given in
Table~\ref{fitall}. The given error bars are formal and should be
treated as lower bounds.

\begin{deluxetable}{ccrcc}
\tablecolumns{5}
\tablewidth{0pt}
\tablecaption{Simultaneous Fit to GD~358 Data Set: \newline 
$\tau_0 = 572.9 \pm  6.1 $~s, $N = 23.5 \pm  0.1$, 
$\theta = 50.5 \pm  0.2^\circ$ }
\tablehead{ \colhead{Period} &\colhead{$\ell$} &\colhead{$m$} & \colhead{Amplitude} &
  \colhead{Phase (rad)} }
\startdata
   962.385 & 1 &  0 &  0.1087 $\pm$  0.0012 &  2.4641 $\pm$  0.0069 \\ 
   852.502 & 1 &  0 &  0.1198 $\pm$  0.0015 &  3.3007 $\pm$  0.0075 \\ 
   810.291 & 1 &  0 &  0.4581 $\pm$  0.0049 &  3.1301 $\pm$  0.0030 \\ 
   575.933 & 1 & -1 &  0.4838 $\pm$  0.0051 &  3.4434 $\pm$  0.0039 \\ 
   574.162 & 1 &  0 &  0.2257 $\pm$  0.0024 &  5.5258 $\pm$  0.0062 \\ 
   573.485 & 1 &  0 &  0.1082 $\pm$  0.0016 &  3.6576 $\pm$  0.0115 \\ 
   571.735 & 1 &  1 &  0.3728 $\pm$  0.0040 &  2.4489 $\pm$  0.0046 \\ 
   465.034 & 1 & -1 &  0.1408 $\pm$  0.0021 &  1.8310 $\pm$  0.0123 \\ 
   464.209 & 1 &  0 &  0.1391 $\pm$  0.0018 &  2.4672 $\pm$  0.0097 \\ 
   463.376 & 1 &  1 &  0.2540 $\pm$  0.0030 &  0.2052 $\pm$  0.0073 \\ 
   423.898 & 1 & -1 &  0.2406 $\pm$  0.0030 &  2.0455 $\pm$  0.0083 \\ 
   422.561 & 1 &  1 &  0.2537 $\pm$  0.0031 &  0.6020 $\pm$  0.0077 
\enddata
\label{fitall}
\end{deluxetable}

To test the sensitivity of the value of $\tau_0$ to the $\ell$
identifications we re-computed fits changing the assumed $\ell$ of the
largest amplitude mode ($P = 810.29$~s) from $\ell=1$ to $\ell=2$.
With this assumption, the best fit resulted for an identification with
$\ell=2$, $m=1$ and had the following parameter values: $ \tau_0 =
569.5 \pm 7.6 $ s, $ N = 14.5 \pm 0.1 $, $ \theta_i = 61.3 \pm
0.1^\circ$. This indicates that the value obtained for $\tau_0$ does
not crucially depend on the mode identifications of each mode in the
fit.

Examination of the fit for each individual run revealed
an apparent modulation of amplitudes, i.e., on some nights the
variations in the light curve were smaller than the fit and on others
they were larger. In addition, looking at data from single sites
suggested that the change from a smaller to a larger amplitude state
alternated on a roughly night to night timescale.

To test whether a geometric effect or a change in the background state
of the star could be causing the night-to-night modulation of the
amplitudes, we went back and re-fit each run in Table~\ref{obslist}
individually. For these fits we fixed the period and amplitude of each
mode to be those given in Table~\ref{fitall} but we allowed the
inclination angle $\theta_i$ and the phases of the modes to vary. We
also allowed $\tau_0$ and $N$ to vary.  Thus, from each run we
obtained a best fit value of the parameters at the time of the run.
In Figure~\ref{theta1} we present the results of this procedure; we
plot the variations in $\theta_i$, $N$, and $\tau_0$ as a function of
time. While no clear trends are seen in $\tau_0$ and $N$ (lower
panels), the variations in $\theta_i$ are suggestive of a periodic
origin (top panel). Assuming a sinusoidal variation, we have included
the best fit sine curve in this plot; it has an amplitude of $14.9 \pm
1.8 ^\circ$ and a period of $2.17 \pm 0.01$ days.  On the other hand,
if we ignore the possible origin of these variations and simply
consider them to be separate measures of $\tau_0$, $N$, and
$\theta_i$, then we obtain fairly conservative limits on the error of
the mean values of these quantities: $\tau_0 = 586.0 \pm 11.7$~s, $N =
27.4 \pm 0.7$, and $\theta_i = 47.5 \pm 2.2^\circ$. We return to the
question of this modulation in section~\ref{obpm}.

\section{The \emph{sforzando}}
 
During the 2006 WET run, GD~358 was stable in that the amplitudes and
phases of its modes were fairly constant over the 3 week length of the
run (except for the modulation detailed above and in
section~\ref{obpm}). However, GD~358 is known to change its pulsation
spectrum on a range of timescales.  A spectacular example of this
behavior occurred in 1996.  Within a period of 36 hrs, all of the
power in the high $k$ range (pulsations with periods $\ga 700$~s)
disappeared within detection limits.  At the same time, GD~358 more
than doubled its apparent pulsation amplitude, with power appearing
almost exclusively at lower $k$, with a period of $\sim 420$ s.
Over the next week, its amplitude decreased to ``normal'' levels,
while the high $k$ power did not return for approximately one month.
This dramatic change is documented in several papers, most recently in
\citet{Kepler03} and \citet{Provencal09}. The episode itself is termed
the ``sforzando'' after a musical term for a sudden and short-lived
increase in loudness.

\begin{figure}
  \centering{\includegraphics[height=1.0\columnwidth,angle=-90]{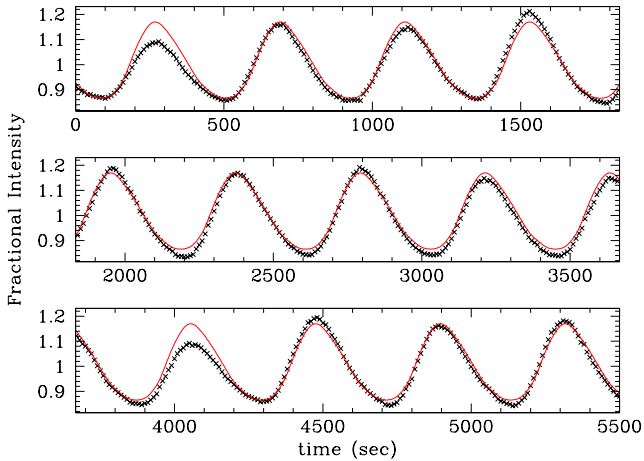}}
  \caption{ A section of the light curve of GD~358 during the
    \emph{sforzando} event in 1996. The crosses are the data points and
    the curve is the best nonlinear light curve fit.}
  \label{whoopsie}
\end{figure}

In Figure~\ref{whoopsie} we show a section of GD~358's light curve
during the \emph{sforzando}. Ironically, during this dramatic episode
the light curve was much simpler. While the amplitude was larger, the
star was essentially a single-mode pulsator and the individual pulses
were more sinusoidal (less nonlinear) than before. Still, GD~358 was
in a transient state and not all pulses were identical. The upper and
lower panels of Figure~\ref{whoopsie} show pulses significantly below
and above the fit, while the middle panel seems to represent a
repeating, ``quasi-static'' state.  Even with these caveats, the
relative simplicity of these data are ideal for nonlinear light
curve fitting.

From a time series analysis of this short section of data, we find
that these pulses have a period of approximately 420.7~s. Assuming
$\ell=1$ and trying all values of $m$, we find the best fit shown in
Figure~\ref{whoopsie}: $m=0$, $\tau_0 =41.6\pm 2.3$~s, $N = 3.6 \pm
0.2$, and $\theta_i = 56.1 \pm 1.1$. For $\ell=1$, $m=1$ a similar
quality fit can be found, but it has a less plausible inclination
angle ($\sim 85^\circ$) that is not consistent with the values of
$\theta_i$ previously found in this paper; even so, $\tau_0=28.1$~s
for this fit.  If we assume $\ell=2$ then the best fit has $\tau_0
\sim 24.2$~s and $m=0$, although it requires such a large intrinsic
amplitude that we consider it unphysical. Summing up, while we have
good reasons for preferring the $\ell=1$, $m=0$ identification for our
fits, the overall value of $\tau_0$ is not strongly dependent on this
identification.

\begin{figure}[t]
  \centering{\includegraphics[height=1.0\columnwidth,angle=-90]{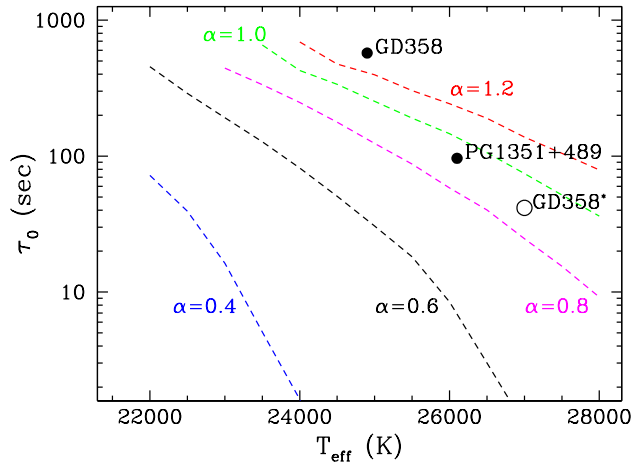}}
  \caption{ A comparison of the derived convective parameters $\tau_0$
    with values expected from ML2/$\alpha$ convection. The labeled
    points are individual objects and the dashed curves are the
    calculations. The label ``GD~358'' represents GD~358 during the
    2006 WET run while ``GD~358$^*$'' stands for GD~358 during its
    \emph{sforzando} episode.}
  \label{DBis}
\end{figure}

Comparing these results to those of section~\ref{lcfits}, we find that
$\tau_0$ was much smaller during the \emph{sforzando} than it was
during the 2006 WET run. This implies that, for whatever reason,
GD~358's convection zone was thinner during the \emph{sforzando}.  As
we can see from Figure~\ref{DBis}, this would imply a $T_{\rm eff}$
several thousand degrees hotter than its normal temperature. For
ML2/$\alpha=1.0$ convection \citep[e.g., ][]{Bohm71}, this would mean
$T_{\rm eff} \sim$~27,000~K, which is two to three thousand degrees
hotter than its normal equilibrium state.

Such an increase in temperature would lead to about a 40\% increase in
the bolometric luminosity. From Figure~\ref{fplot} we see that such an
increase in the bolometric luminosity translates into a 15--20\%
increase in intensity as measured in a passband centered at 5200~\AA.
While such a passband is reasonable for a CCD $+$ BG40/S8612 filter $+$
atmosphere, the observations in 1996 were made with phototubes, and
these are much more blue sensitive. Thus, for a phototube passband
centered around 3800~\AA\ we expect intensity increases of about
20--25\%.

\begin{figure}[t]
  \centering{\includegraphics[height=1.0\columnwidth,angle=-90]{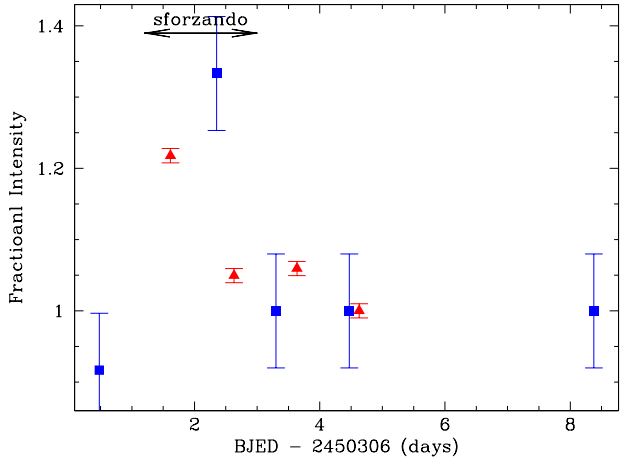}}
  \caption{ The relative intensity of GD~358 as measured relative 
    to its levels after the \emph{sforzando} episode. The triangles
    are data from McDonald Observatory and the squares are data from
    Mt.~Suhora Observatory. Both data sets indicate a jump in
    intensity near BJED $\sim 2450311$.  }
  \label{wamps}
\end{figure}

As reported in Table~5 of \citet{Provencal09}, such an increase in
intensity was observed in GD~358 during the \emph{sforzando} event. In
Figure~\ref{wamps} we plot these data, where the triangles are data
from McDonald Observatory and the squares are data from Mt.~Suhora
Observatory in Poland. Both curves are normalized to a value of one
near BJED $\sim 2450311$. We see that these data, although sparse,
strongly suggest that GD~358 underwent a sharp intensity increase and
decline of about 20--30\% during this event.  This corroborates our
earlier interpretation that GD~358 had a thinner convection zone with
a smaller value of $\tau_0$, and that this thinness was due to its
surface layers being temporarily hotter.

We note that similar results and conclusions have previously been made
by \citet{Weidner03}. Using the approach of \citet{Ising01} they were
the first to make numerical simulations of GD~358's light curve during
the \emph{sforzando} event. As in our present analysis, they found that the
shape of the light curve could only be reproduced for higher effective
temperatures than normally assumed for this star, suggesting a value
of $\teff \sim~$27,000~K.

\subsection{The Amplitude of the $k=8$ Mode}

The physical origin of the \emph{sforzando} event is completely
unknown.  All we know for certain is that the amplitudes of the
high-$k$ modes rapidly decreased, leaving power in the $k=8$ mode,
which itself \emph{increased} in amplitude by approximately a factor
of 30. While we have no model to explain the disappearance of power in
the high-$k$ modes, in this section we consider what would happen to
the apparent amplitude of the $k=8$ mode if the convection zone were
suddenly removed.

One effect of the thermal response of the convection zone is to
attenuate the amplitude of the flux variations incident on its base.
It acts as a low-pass filter, reducing the photospheric amplitude of a
mode according to
\begin{equation}
  \frac{\Delta F_{\rm phot}}{F_0} = \frac{1}{\sqrt{1+(\omega \tau_0)^2}} \,
  \frac{\Delta F_{\rm base}}{F_0},
  \label{amps}
\end{equation}
where $F_0$ is the equilibrium value of the flux, $F_{\rm base}$ and
$F_{\rm phot}$ are the instantaneous fluxes at the base of the
convection zone and at the photosphere, respectively, and $\omega$ is
the angular frequency of the mode \citep{Goldreich99a,Wu99}.

If we assume that the amplitude of the $k=8$ mode at the base of the
convection zone was constant throughout the \emph{sforzando},
equation~\ref{amps} shows that a decrease in $\tau_0$ would naturally
lead to an increase in observed amplitude. Thus, for a 420~s mode we
would see an increase of approximately a factor of 8 in its apparent
amplitude as $\tau_0$ goes from $\sim 600$~s to $\sim 40$~s. This is a
large factor, but it is still much less than the factor of $\sim 30$
increase that was actually observed. Thus, the $k=8$ appears to have
increased its intrinsic amplitude by of order a factor of 4 during the
\emph{sforzando}. How it was able to do this on so short a time scale
remains a mystery.

\section{The Convective Time Scale $\tau_0$}

In Figure~\ref{DBis} we plot the known determinations of the
convective response timescale, $\tau_0$, versus \teff\ for the DBVs.
Based on the discussion in the previous section, GD~358 is plotted
twice: once for the 2006 data and once for the 1996 data. In addition,
we show the position of the DBV PG1351+489 \citep[$\tau_0 \sim
100$~s,][]{Montgomery05a}, where we have assumed the pure He fit for
its \teff\ \citep{Beauchamp99}. As expected, the data indicate an
increase in the depth/mass of the convection zone with decreasing
\teff.

We also plot lines in Figure~\ref{DBis} showing the predictions of
ML2/$\alpha$ convection \citep{Bohm71} for various values of $\alpha$.
We see that while low values of $\alpha$ are excluded ($\alpha \la
0.6$), values in the range 1.0 to 1.2 provide a reasonable description
of how $\tau_0$ varies with \teff. We note that the $\log g$ values
determined for these stars are nearly identical, so they \emph{do}
form an actual sequence in \teff. In general, however, $\tau_0$ is
also a function of $\log g$, albeit a somewhat weaker one. 

Our ultimate goal is to map $\tau_0$ as a function of both \teff\ and
$\log g$ for both the DBV and DAV instability strips. This will
provide important reference points for new hydrodynamic simulations of
convection which are starting to come online
\citep[e.g.][]{Muthsam09}. For instance, for a given white dwarf one
can use the measured \teff\ and $\log g$ values and perform a
hydrodynamic simulation of its convection zone. Then, using
equations~3 and 5 of \citet{Wu01} one can compute $\tau_0$. This value
of $\tau_0$ can then be compared with the value derived from the light
curve fits outlined above.

\section{The Oblique Pulsator Model}
\label{obpm}
\subsection{The Formalism}
\label{obpmf}

In section~\ref{lcfits} we found evidence of a modulation of the
inclination angle $\theta_i$ (see Figure~\ref{theta1}).  This
modulation has a formal significance level of $14.9/1.8 \sim 8\,
\sigma$, which cries out for a physical interpretation. The most
obvious is some form of the ``oblique pulsator'' model, in which the
pulsation axis is inclined with respect to the rotation axis, and this
axis precesses as the star rotates \citep{Kurtz86,Kurtz82}. A magnetic
field that is inclined to the rotation axis is usually invoked, and
the pulsations are assumed to be aligned with the magnetic axis.

\begin{deluxetable*}{ccccc}
\tablecolumns{5}
\tablewidth{0pt}
\tablecaption{Frequency solution for oblique pulsation model: \newline 
  $f_{\rm rot}=5.362 \pm 0.003$~$\mu$Hz
  \label{obp}}
\tablehead{\colhead{Frequency ($\mu$Hz)} & \colhead{Amplitude (mma)} &
 \colhead{Phase (rad)} & \colhead{$\Delta \Phi/2 \pi$} 
 & \colhead{Independent Freqs.\tablenotemark{a}}
}
\startdata
\multicolumn{5}{c}{\bf Triplet 1} \\[0.4em]
\hline\\[-0.5em]
 1736.302 $\pm$  0.004 &  16.77 $\pm$  0.13 &  0.739 $\pm$  0.008 & & 1736.302 $\pm$  0.001 \\
 1741.664 $\pm$  0.003 &  10.81 $\pm$  0.13 &  2.832 $\pm$  0.012 & 0.510 $\pm$  0.013 & 1741.665 $\pm$  0.001 \\
 1747.027 $\pm$  0.004 &  \phn 1.75 $\pm$  0.13 &  1.724 $\pm$  0.075 & & 1746.673 $\pm$  0.007 \\[-0.5em]
\cutinhead{\bf Triplet 2}
\\[-0.7em]
 1738.362 $\pm$  0.005 & \phn 0.95 $\pm$  0.13 &  1.268 $\pm$  0.138 & & 1737.962 $\pm$  0.007 \\
 1743.725 $\pm$  0.003 & \phn 5.56 $\pm$  0.13 &  0.174 $\pm$  0.023 &  0.517 $\pm$ 0.023 & 1743.738 $\pm$  0.002 \\
 1749.087 $\pm$  0.005 &     12.55 $\pm$  0.13 &  2.117 $\pm$  0.011 & & 1749.083 $\pm$  0.001 \\[-1.0em]
\enddata
\tablenotetext{a}{These are the unconstrained frequency fits of
  \citet{Provencal09}.}
\end{deluxetable*}

While this hypothesis does introduce several unknowns (a magnetic
field, the angle between the magnetic and rotation axes, etc.) it also
makes three testable predictions. If we take $f$ and $f_{\rm rot}$ to
be the mode frequency in the frame of the star and the rotational
frequency, respectively, then an $\ell=1$ mode with frequency $f$
aligned with the magnetic axis will appear as three separate peaks in
the Fourier transform of the light curve, with frequencies of
$f-f_{\rm rot}$, $f$, and $f+f_{\rm rot}$. For clarity we will refer
to these peaks as ``geometric'' peaks as they only appear because the
pulsation axis spins around the rotation axis, leading to a periodic
apparent amplitude modulation of the mode. This amplitude modulation
manifests itself in the Fourier transform as two additional
``geometric'' peaks on either side of the original frequency, with the
beating of these three peaks producing the periodic amplitude changes.
As this splitting is caused by the rotation of the star, these
frequencies \emph{must} be equally split to within measurement errors.
This is the first and most stringent condition the model must face.

Second, the oblique pulsator model predicts a specific phase
relationship for the peaks in a given geometric triplet (or for higher
$\ell$, a $(2\ell+1)$-multiplet). From \citet{Kurtz86}, an $\ell=1$ mode 
generically has luminosity variations given by
\begin{eqnarray}
  \Delta L/L & = &
  A_- \cos[(\omega-\Omega)t + \phi] \nonumber \\
  & + & A_0 \cos[\omega t + \phi]
  +A_+ \cos[(\omega+\Omega)t + \phi],
\end{eqnarray}
where $\omega$ and $\Omega$ are $2 \pi$ times $f$ and $f_{\rm rot}$,
respectively (see Appendix~\ref{app1} for the complete expressions).
These components will only have the same phase $\phi$ for a particular
choice of the zero point of time. For other zero points one can show
that this phase relation translates to
\begin{equation}
  2 \Phi_0 - \left(\Phi_+ + \Phi_-\right) = 0,
  \label{phi1}
\end{equation}
where $\{\Phi_-, \Phi_0,\Phi_+\}$ are the measured phases of the
respective components. In general, the product $A_- \times A_+$ can be
negative. For this case, if we define the amplitudes always to be 
positive and absorb any minus signs into the phase for each peak, the
relation becomes
\begin{equation}
  2 \Phi_0 - \left(\Phi_+ + \Phi_-\right) = \pi.
  \label{phi2}
\end{equation}
Defining $\Delta \Phi \equiv 2 \Phi_0 -(\Phi_++\Phi_-)$, then from the
sign of the amplitudes in equations~(\ref{ampf})--(\ref{ampl}) we see
that $\Delta \Phi/2 \pi = 0$ for $m=0$ modes and $\Delta \Phi/2 \pi =
0.5$ for $m=\pm 1$ modes.  These phase relations, while less iron-clad
than the equal spacing of the triplets, should be satisfied within the
errors for geometric peaks split by oblique rotation.

A third condition/prediction is the relative amplitudes of the
geometric peaks. Such a calculation makes assumptions concerning the
nature and strength of the magnetic field, so this prediction of the
model is the least reliable of the three. In
equations~(\ref{ampf})--(\ref{ampl}) in Appendix~\ref{app1} we give
expressions for the amplitudes of the various components of $\ell=1$
modes perturbed by a magnetic field and oblique rotation. The relevant
parameters are the inclination angle of the rotation axis,
$\theta_i$, the obliquity of the magnetic axis, $\beta$, and $x_1$.
The parameter $x_1$ is given by
\begin{equation}
  x_1 \equiv \frac{\omega^{(1)}_0 - \omega^{(1)}_1}{C_{k\, \ell} \,\,\Omega},
  \label{xdef}
\end{equation}
where $C_{k\, \ell}$ is the rotational splitting coefficient due to
the Coriolis force, $\Omega$ is the angular frequency of rotation, and
$\omega^{(1)}_0$ and $\omega^{(1)}_1$ are the perturbations to the
frequencies of the $m=0$ and $m=1$ \emph{intrinsic} modes due to the
magnetic field, respectively. Even though this is clearly a more model
dependent statement, we would still hope that the amplitudes could
approximately be fit and/or predicted within the oblique pulsator
formalism.

\subsection{The Results}

First, we consider a fit to the $k=12$ region of the FT, from
1730--1750~$\mu$Hz.  \citet{Provencal09} found 6 significant peaks in
this region.  We interpret these peaks as resulting from 2 components
of an $\ell = 1$ mode, each split into a geometric triplet by oblique
pulsation.  These two original peaks are part of an intrinsic triplet
produced by standard rotational splitting, but the amplitude of the
third member is below our detection threshold. 

\begin{figure}[t]
  \centering{\includegraphics[height=1.0\columnwidth,angle=0]{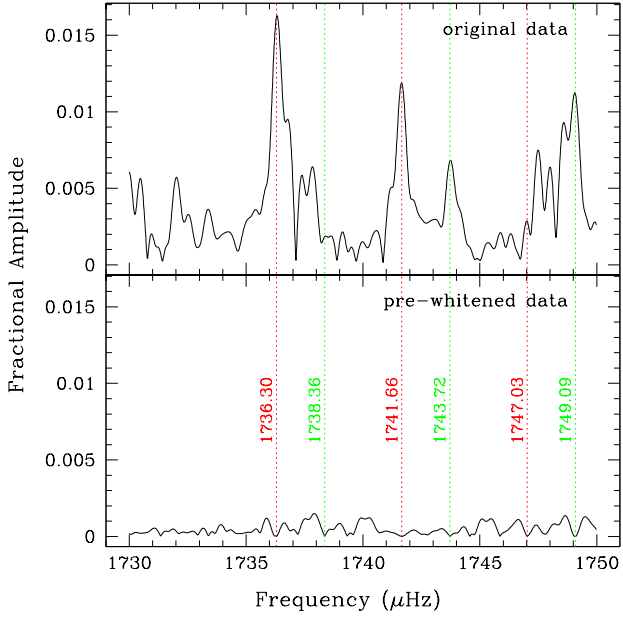}}
  \caption{The effect of pre-whitening by two identically split
    triplets in the oblique pulsator model. The top panel is the
    original FT of the $k=12$ region and the lower panel is the result
    of pre-whitening by the two pairs of triplets. The splitting within
  each triplet is $5.362 \pm 0.003 \mu$Hz.}
  \label{prewhiten}
\end{figure}

We fit two sets of exactly evenly split triplets to the data set,
where the central frequencies of each triplet are 1741.664 and 1743.725
$\mu$Hz, and the value of the splitting is 5.362~$\mu$Hz. The values
of the fit parameters we obtained are given in Table~\ref{obp}.  In
Figure~\ref{prewhiten} we show the result of pre-whitening the $k=12$
region by this solution. The reduction in power of the FT is very
significant, showing that evenly split triplets are a good
representation of the data. This is a necessary condition for the
oblique pulsator model to be applicable.

\begin{figure}[b]
  \centering{\includegraphics[height=1.0\columnwidth,angle=-90]{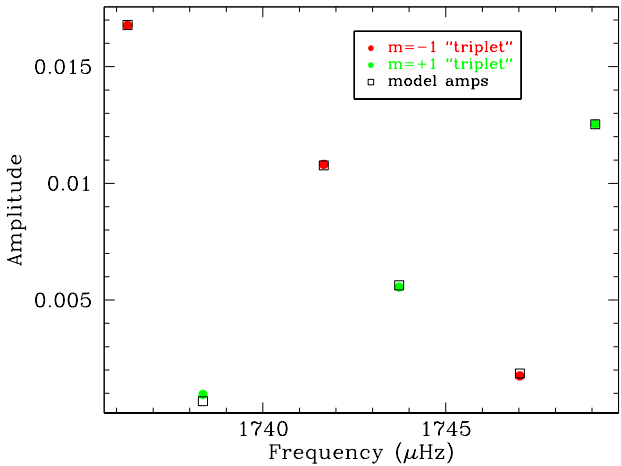}}
  \caption{A fit of the calculated amplitudes (open squares) to the
    observed amplitudes (points) for the different components of
    ``k=12'' assuming the oblique pulsator model. The errors on the
    observed amplitudes are smaller than the size of the points.}
  \label{amp}
\end{figure}

Second, we consider the phase relations within each of these equally
split triplets. From Table~\ref{obp} we see that the first triplet has
$\Delta \Phi/2\pi = 0.510 \pm 0.013$ and that the second triplet has
$\Delta \Phi/2\pi = 0.517 \pm 0.023$. The oblique pulsator model also
passes this test with flying colors. In addition, the fact that
$\Delta \Phi/2\pi \sim 0.5$ rather than 0.0 implies that the modes are
not axi-symmetric, i.e., $|m|=1$ for both modes. For whatever reason,
the $m=0$ member of the original triplet is not present at observable
amplitudes.

Finally, we wish to test the ability of the oblique pulsator model to
adequately reproduce the amplitudes of the geometric peaks. To
calculate the amplitudes we use the analytical expressions given in
Appendix~\ref{app1} \citep[see][]{Unno89,Kurtz86} and to perform the
fits we have used a genetic algorithm \citep{Charbonneau95}.
This allows us to search the $m$ values of each intrinsic triplet as
well as the values of the parameters $\theta_i$, $\beta$, and $x_1$.
As we demonstrate below, such a fit is more constrained than one would
think given the six data points and five free parameters.

In Figure~\ref{amp} we show the result of the fit to the amplitudes in
Table~\ref{obp}. The fit is quite impressive. It has the added bonus
that triplets~1 and 2 are \emph{required} to originate from $|m|=1$
intrinsic modes and they must have opposite signs. This corroborates
our earlier result based on the phases that both triplets originated
from $|m|=1$ intrinsic modes.

Given the near equality of the number of data points and free
parameters we wished to assess how easily our model could reproduce
any data set.  To do this, we randomly generated amplitudes for 500
pairs of triplets and fit them in the same way we fit the data,
normalizing the residuals by the mean squared amplitudes of the two
triplets. We found that in only 11 cases out of 500 were the random
amplitudes better fit by our model than the measured amplitudes were.
Thus, we conclude that our amplitude fits are significant at the $\sim
98$\% level.

\begin{figure}[t]
  \centering{\includegraphics[height=1.0\columnwidth,angle=0]{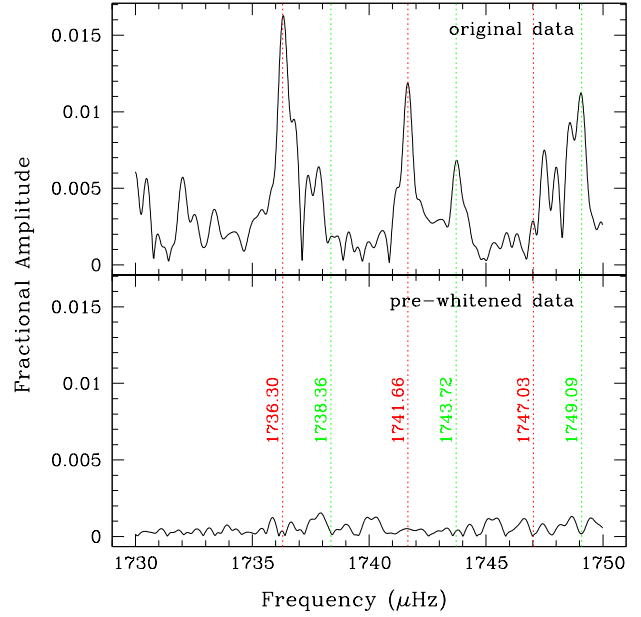}}
  \caption{Pre-whitening of the same region as in Figure~\ref{prewhiten}
    but using the full oblique pulsator model for the amplitudes,
    frequencies, and phases. Thus, not only are the frequency
    splittings constrained, but the amplitudes within a triplet and
    their relative phases are as well.  }
  \label{prewhiten2}
\end{figure}

As a final check on this procedure, we use these amplitudes together
with equations~\ref{ampf} and \ref{ampl} to pre-whiten the $k=12$
region of the FT, as shown in Figure~\ref{prewhiten2}.  Although
considerably more constrained, this pre-whitening is clearly just as
effective as that shown in Figure~\ref{prewhiten}. Thus, the oblique
pulsator model passes all tests with flying colors when applied to the
$k=12$ region of the FT.

\subsection{Interpretation of the Oblique Pulsator Fits}

The relative amplitudes of the peaks within a triplet split by oblique
pulsation/rotation depend on $\theta_i$, $\beta$, and $x_1$ (see
section~\ref{obpmf} for a definition of these quantities).  From the
amplitude fits obtained in the previous section, we find that
$\theta_i=46.3^\circ$, $\beta=31.8^\circ$, and $x_1=5.65$. This value
of $\theta_i$ is close to that obtained from our nonlinear light curve
fits, $47.5\pm 2.2^\circ$ (section~\ref{lcfits}). This provides an
important consistency check on both methods.

Given that $\theta_i$ is the angle between the rotation axis and our
line-of-sight and that $\beta$ is angle between the rotation and
pulsation/magnetic axes, the effective inclination angle of a mode
aligned with the magnetic axis should vary between $\theta_i-\beta$
and $\theta_i+\beta$ as the star rotates, i.e., between $14.5^\circ$
and $78.1^\circ$.  From the top panel of Figure~\ref{theta1} we see
that $\theta_i$ varies with about half of this amplitude. The
straightforward explanation for this is that only the $k=12$ modes are
affected by oblique rotation, while the other modes are aligned with
the rotation axis. As a result, the value of $\theta_i$ shown in
Figure~\ref{theta1} represents the average inclination angle for all
modes. Averaging over all modes yields an inclination angle which
varies by about $15^\circ$ rather than the full $32^\circ$ experienced
by the $k=12$ modes.

Finally, the parameter $x_1$ is a measure of the relative strength of
the Coriolis and magnetic perturbations; the fit value $x_1=5.65$
indicates that the magnetic field is the dominant perturber. This
reinforces the idea that the pulsations of these modes are tied to the
magnetic axis. In addition, for a large-scale dipole field simple
perturbation theory applied to $\ell=1$ and 2 modes yields
\begin{equation}
 \omega^{(1)}_{|m|} \propto \left\{\begin{array}{cl}
                \frac{2}{5} + \frac{2}{5} m^2, & \ell = 1\\[.5em]
                \frac{18}{7} - \frac{2}{7} m^2, & \ell = 2
         \end{array}
         ,
         \right.
\label{alm}
\end{equation}
where $\omega_{|m|}^{(1)}$ is the perturbation to the frequency of the
mode due to the magnetic field \citep{Montgomery94,Jones89}.  Together
with equation~(\ref{xdef}), the fact that $x_1$ is positive means that
$\omega_0^{(1)} > \omega_1^{(1)}$, which in turn suggests that $\ell
\ge 2$. In reality, this indicates that these modes, at least in the
surface layers, may not be pure $\ell=1$ modes but contain a mixture
of other higher $\ell$ components.

The success of the oblique pulsator model in explaining the $k=12$
peaks constitutes evidence that oblique pulsation plays an important
role in GD~358. However, there is little conclusive evidence that other
modes in this star are undergoing oblique pulsation. On the contrary,
the stable $k=8$ and 9 triplets are consistent only with the
traditional case of the pulsation axis aligned with the rotation axis.
A possible explanation is that the outer turning points for the $k=8$
and 9 modes are farther from the surface than they are for the $k=12$
modes, so if the dominant magnetic effects are also confined to the
surface layers then the lower $k$ modes would not be affected by the
magnetic field and the higher $k$ modes would.

The \emph{magnitudes} of the multiplet splittings/fine structure also
present a challenge.  For instance, the oblique pulsator model
requires a rotation frequency of 5.36~$\mu$Hz, whereas the splittings
of the $k=8$ and 9 modes, believed to result from standard rotational
splitting, give a value of 7.5~$\mu$Hz \citep{Provencal09,Winget94}.
Since different modes sample different regions of the star's interior,
differential rotation in the radial direction could explain these
results.

A further concern is the internal consistency of the oblique pulsator
fit. The central components of each of the triplets in Table~\ref{obp}
should themselves be members of an intrinsic triplet which is split by
rotation. This splitting is observed to be $2.06~\mu$Hz and should
equal $2 C_{k\,\ell} f_{\rm rot} \cos \beta$ for solid body rotation
of the star. This is satisfied if $C_{k\,\ell}\sim 0.225$ instead of
$C_{k\,\ell} \sim 0.5$ as is expected for $\ell=1$ modes. The
splitting within a geometric triplet is a direct measure of the
\emph{surface} rotation rate, while the difference in frequency of the
central components of these triplets (the ``intrinsic'' modes) is
given by $2 C_{k\,\ell} f_{\rm rot} \cos \beta$, where $f_{\rm rot}$
represents a \emph{radial average} of the rotation profile. Thus,
differential rotation, with the interior rotating more slowly than the
surface layers, could explain these values.  An alternate explanation
is that the magnetic field mixes higher $\ell$ components into the
spatial structure of the $k=12$ modes. They would therefore have
rotational constants $C_{k\,\ell}$ indicative of higher $\ell$ modes.
Since $C_{k\,\ell} \sim 1/\ell (\ell+1)$ for moderate to high $k$
modes, these modes would have correspondingly smaller values of
$C_{k\,\ell}$.

A further possibility is that the intrinsic modes have $\ell=2$
instead of $\ell=1$. In this case only 3 of the 5 possible geometric
peaks would be large enough to be detectable. First, we note that the
predicted phase relations are the same as for the $\ell=1$ case, so
the measured phase relations support either case equally.  Also, since
$C_{k\,\ell} \sim 1/6$ for $\ell=2$, using the rotational splitting of
the $k=8$ and 9 modes yields a splitting of $2 C_{k\,\ell} f_{\rm rot}
\cos \beta \sim 2.2\,\mu$Hz, which is close to the measured value of
$2.06~\mu$Hz. In this case, though, the value of the inclination
angle, $\theta_{i}=22.1^\circ$, does not agree with that from the
light curve fits. Also, the $\ell=2$ fits are somewhat less
constrained than the $\ell=1$ fits: fitting 500 random amplitudes in
the same way as the data, 12\% of the fits were better than the fit to
the data, making the $\ell=2$ fit significant at the 88\% level.

Even so, given the ability of the oblique pulsator model to describe
(1) the power in the $k=12$ region, (2) the phase relations between
the geometric triplets, and (3) the relative amplitudes within the
triplets (to some extent), we believe that the above-mentioned
inconsistencies do not warrant rejection of the model but rather point
us in the direction of future improvements. For instance, the
amplitude calculations include magnetic effects in a fairly simple
way; more sophisticated treatments may be necessary to adequately
model the observations. Also, differential rotation in the radial
direction may be able to resolve the seeming discrepancy of the
$C_{k,\ell}$ values for the $\ell=1$ case.

\section{Discussion}

In many respects GD~358 is a ``simple'' star to analyze. For instance,
the spacing between the periods of its multiplets suggests that we are
looking at successive radial overtone numbers of $\ell=1$ modes. This
is based both on the fact that (1) given its distance, mass, and
luminosity, only $\ell=1$ modes allow a consistent solution
\citep{Bradley94a} and (2) in many cases these multiplets are well
defined triplets \citep{Winget94a}. Furthermore, as a DBV it has no
hydrogen layer, so fewer parameters are needed to model its structure.
Indeed, it is the best studied white dwarf variable and significant
constraints have been placed on its interior structure
\citep{Metcalfe00,Metcalfe03}.

On the other hand, GD~358 is in many respects a complicated star to
model.  For instance, while the amplitudes of its modes can be fairly
constant during a WET run, over timescales of a few months the
amplitudes can change significantly. The most extreme case of
amplitude change is the previously discussed \emph{sforzando}, in
which dramatic amplitude changes occurred on a timescale of a day or
less. We currently have no theory that adequately describes these
changes.

What we \emph{do} know is that two independent lines of evidence
indicate that GD~358 was hotter during the \emph{sforzando} event.
The first is that the average flux of GD~358 relative to the
comparison stars is larger, and the second is that the nonlinear light
curve fits indicate a thinner convection zone---presumably the result
of an increase in surface temperature. If we assume that GD~358
remained hotter for approximately a day then such a temperature
increase would require a total energy input of approximately $6\times
10^{36}$~ergs. While detailed nonadiabatic models of GD~358's
pulsational state just prior to the \emph{sforzando} would be
required, rough estimates indicate that this amount of energy may
typically be present in the higher $k$ modes. Thus, if these modes
were somehow damped and deposited their pulsation energy in the
surface layers of the star over the period of a day, this would
explain GD~358's temporary temperature increase. Such a scenario would
explain both the temperature increase and the disappearance of the
high $k$ modes, although more detailed models would be necessary to
check quantitatively the energetics.

We note that an increase in the apparent amplitude of the $k=8$ mode
is a generic feature of a thinner convection zone (e.g., smaller value
of $\tau_0$), although the predicted factor ($\sim 8$) is still less
than what was observed ($\sim 30$). While speculative, it is also
possible that some mechanism may have transferred power from the high
$k$ to the low $k$ modes, although we currently do not understand how
this would proceed. Perhaps least likely is the possibility that the
hotter ``equilibrium'' state of the \emph{sforzando} allowed mode
growth and damping to occur on timescales given by linear theory and
that these timescales were many orders of magnitude shorter than
expected.

Another possibility is that the event was tied to a magnetic
phenomenon and that this phenomenon somehow led to a temperature
increase as well as a change in the amplitudes of the various modes.
There are several reasons that this is not completely far-fetched.
First, the oblique pulsator model used here requires a mechanism to
align modes with an axis other than the rotation axis, and a magnetic
field can provide this. Second, the $k=8$ and 9 triplets are
asymmetric in the sense that the $m=0$ mode is closer to either the
$m=+1$ or $m=-1$ mode \citep[see Figures~18 and 19 of][]{Provencal09}.
Perhaps not coincidentally, the asymmetry seen in the splittings of
the $k=8$ and 9 modes reversed themselves at the time of the
\emph{sforzando}, and this change has persisted ever since.  This
timescale of years is consistent with what is seen in the magnetic
cycle of the Sun and other stars \citep{Elsworth90,Libbrecht90}.
Finally, small but definite shifts in frequency of order 0.5~$\mu$Hz
have been seen on timescales as short as 3 weeks. These could be a
sign of magnetic activity: changes in the surface magnetic field could
produce slight perturbations in the mode frequencies \citep{Jones89},
and the timescale of a few weeks for these to take place again seems
plausible.

A final unresolved issue for GD~358 is differential rotation.  Taken
at face value, the difference in triplet splitting between high $k$
and low $k$ modes implies significant differential rotation
\citep{Winget94}.  In addition, the oblique pulsator model as applied
in this paper implies differential rotation: the $k=12$ region
requires a surface rotation rate of $\sim 2.17$~days whereas the $k=8$
and 9 splittings give a rotation rate of $\sim 1.5$~days.  The $k=8$
and 9 splittings reflect a bulk average of the rotation rate whereas
the frequency differences within the geometric triplets in the $k=12$
region give us the rate at the surface. In addition, the frequency
splitting between the intrinsic modes in the $k=12$ region also
implies differential rotation, albeit with the interior rotating less
rapidly than the surface. Previously, detailed examinations of the
frequency splittings in GD~358 as a function of $k$ have been
inconclusive \citep{Kawaler99}, but such analyses did not take into
account the possibility of oblique pulsation.
      
An intriguing possibility is that the pulsations themselves lead to
differential rotation. \citet{Townsend09} has recently shown that
g-modes in massive stars can transport angular momentum relatively
rapidly compared to evolutionary timescales.  This transport occurs
predominantly in regions in which the mode is driven and/or damped and
leads to differential rotation. If this effect occurs in white dwarfs
then it could alter the rotation rate from the surface down past the
base of the convection zone into the radiative damping layers.  If the
rotation profile of these outer layers is continually changing then
this would also explain the small shifts in frequency which have been
detected in GD~358. In addition, a more dramatic shift in the rotation
profile could be associated with the \emph{sforzando} event.  Perhaps
the outer part of GD~358's rotation profile experienced a shift during
this event and it has persisted in its new state for the past several
years. This would explain the shift in asymmetry of the $k=8$ and 9
triplets that also occurred at this time \citep[see][]{Provencal09}.

\newpage 

\section{Conclusions}

In this paper we have extended our nonlinear light curve fitting
technique to the multiperiodic pulsator GD~358. Our fit to the 2006
WET data provides a good match to the light curves and we find that
the thermal response time of its convection zone is $\tau_0 = 572.9
\pm 6.1$~s. This is considerably larger than that of the star
PG1351+489, for which $\tau_0 \sim 100$~s \citep{Montgomery05a}.  This
difference in $\tau_0$ is consistent with the effective temperatures
of these stars: the pure He solution for PG1351+489 yields a \teff\
which is $\sim 2,000$~K hotter than that for GD~358.

We also obtained a fit to the light curve of GD~358 during the
\emph{sforzando} event in 1996. These fits showed that GD~358 had
$\tau_0 \sim 42 \pm 2$~s, a value much less than that determined from
the 2006 data. This suggests that its effective temperature was
approximately 2,000~K hotter in 1996 than in 2006, and this is
consistent with the estimate of \citet{Weidner03} that the light curve
shape suggests $\teff \sim 27,000$~K. Independent evidence of GD~358's
brightness relative to comparison stars is also consistent with such a
temperature increase at the time of the \emph{sforzando}
\citep{Provencal09}. The physical origin of this temperature increase
will be the subject of future work.

As expected, these data indicate an increase in the depth/mass of the
convection zone with decreasing \teff. A similar trend is given by
ML2/$\alpha = 1.1$ convection \citep{Bohm71}, although the slope of
the theoretical relation appears less steep than that of the data. In
addition, lower values of $\alpha \la 0.6$ are excluded.  In
general, $\tau_0$ is also a function of $\log g$, albeit a somewhat
weaker one.  Our ultimate goal is to map $\tau_0$ as a function of
both \teff\ and $\log g$ for both the DBV and DAV instability strips.
These data will provide insight into the physics of convection, still
one of the largest uncertainties in stellar modeling. They will also
serve as important constraints for new hydrodynamic simulations of
convection which are starting to come online.

Multiple lines of evidence point to some of GD~358's modes undergoing
oblique pulsation, in particular, the peaks in the $k=12$ region.
First, these peaks can be fit with two sets of exactly evenly spaced
triplets. Second, the relative phases of each of the components within
the triplet indicate that each originates as a single $m=-1$ or +1
mode aligned with the magnetic axis; as the star rotates, the magnetic
axis precesses around the rotation axis, generating a triplet for each
intrinsic mode. Finally, the oblique pulsator model qualitatively and
quantitatively fits the amplitudes of the peaks seen in the Fourier
transform. Taken together, this marks the first time that oblique
pulsation has been seen in a white dwarf variable.

Having now identified the characteristics of oblique pulsation in
GD~358 we now know what to look for in other white dwarf variables; we
have found preliminary indications of it in other stars and in other
data sets of GD~358. As discussed in the previous sections, oblique
pulsation may prove to be a diagnostic of both the magnetic field and
its changes as well as a diagnostic of differential rotation. This
opens an exciting chapter in the seismology of these objects.

\acknowledgments

This research was supported in part by the Delaware Asteroseismic
Research Center, the National Science Foundation under grant
AST-0909107, and the Norman Hackerman Advanced Research Program under
grant 003658-0255-2007.  AJS thanks the National Science Foundation
for support under grant AST-0607840. The Delaware Asteroseismic
Research Center (DARC) is grateful for the support of the Crystal
Trust Foundation and Mt.\ Cuba Observatory.  DARC also acknowledges
the support of the University of Delaware, through their participation
in the SMARTS consortium.

\appendix

\section{Equations of Oblique Pulsation for $\ell=1$ and 2 modes}
\label{app1}

The formulae given below are taken from \citet{Unno89}. Following
\citet{Kurtz86} we use $\omega$ instead of $\sigma$ for the angular
frequencies of the modes.  We use $\omega^{(0)}$ for the unperturbed
frequency of the mode in a non-rotating, non-magnetic star, and
$\omega_{|m|}^{(1)}$ for the perturbation to this frequency due to the
magnetic field (but \emph{not} due to rotation). The pulsation axis is
assumed to be aligned with the magnetic axis, which makes an angle of
$\beta$ with the rotation axis, and $\theta_i$ is taken to be the
angle between the rotation axis and our line-of-sight. As the
pulsation and magnetic axes rotate around the star, we find that a
mode of given frequency and $\{\ell,m\}$ values is split into a
triplet of peaks. We tabulate below the resulting time dependence of
modes having the given values of $\ell$ and $m$.

In the following, $x_1$ is defined to be
\begin{equation}
  x_1 \equiv \frac{\omega^{(1)}_0 - \omega^{(1)}_1}{C \,\,\Omega},
\end{equation}
where $C$ is the rotational splitting coefficient due to the Coriolis
force for $\ell=1$ modes and $\Omega$ is the angular velocity of
rotation at the stellar surface. Since $x_1$ is the ratio of the
rotational splitting to the splitting induced by the magnetic field it
provides a useful estimate of the relative importance of the two
effects. Also, since it depends on the difference in magnetic
splitting between $m=0$ and $|m|=1$ modes, it is sensitive to the
geometry of the magnetic field. With these definitions, the luminosity
perturbations associated with $\ell=1$ oblique pulsation are given
below:
\\


$\ell=1$, $m=-1$:
\begin{eqnarray}
  \Delta L/L  & = &
  \frac{1}{\sqrt{2}} \left(1 - \frac{1+\cos\beta}{x_1}\right)
  \sin^2\frac{\beta}{2} \,\,\sin \theta_i \,\,
  \cos\left[\left(\omega^{(0)} + \omega_1^{(1)}-C\,\Omega \cos \beta - \Omega\right) t + \phi\right] 
  \nonumber \\
  & + & 
  \frac{1}{\sqrt{2}} \left(1 - \frac{\cos\beta}{x_1} \right)
  \sin\beta \,\,\cos \theta_i \,\,
  \cos\left[\left(\omega^{(0)} + \omega_1^{(1)}-C\,\Omega \cos \beta\right) t + \phi\right]
  \nonumber \\
  & - & 
  \frac{1}{\sqrt{2}} \left(1 + \frac{1-\cos\beta}{x_1} \right)
  \cos^2\frac{\beta}{2} \,\,\sin \theta_i \,\,
  \cos\left[\left(\omega^{(0)} + \omega_1^{(1)}-C\,\Omega \cos \beta + \Omega\right) t + \phi\right]
  \label{ampf}
\end{eqnarray}

$\ell=1$, $m=0$:
\begin{eqnarray}
  \Delta L/L  & = &
  \frac{1}{2} \left(1 + \frac{1}{x_1}\right)
  \sin\beta \,\,\sin \theta_i \,\,
  \cos\left[\left(\omega^{(0)} + \omega_0^{(1)} - \Omega\right) t + \phi\right] 
  \nonumber \\
  & + & 
  \cos\beta \,\,\cos \theta_i \,\,
  \cos\left[\left(\omega^{(0)} + \omega_0^{(1)}\right) t + \phi\right]
  \nonumber \\
  & + & 
  \frac{1}{2} \left(1 - \frac{1}{x_1} \right)
  \sin\beta \,\,\sin \theta_i \,\,
  \cos\left[\left(\omega^{(0)} + \omega_0^{(1)}+\Omega\right) t + \phi\right]
\end{eqnarray}

$\ell=1$, $m=1$:
\begin{eqnarray}
  \Delta L/L  & = &
  \frac{1}{\sqrt{2}} \left(1 - \frac{1-\cos\beta}{x_1} \right)
  \cos^2\frac{\beta}{2} \,\,\sin \theta_i \,\,
  \cos\left[\left(\omega^{(0)} + \omega_1^{(1)}+C\,\Omega \cos \beta - \Omega\right) t + \phi\right] 
  \nonumber \\
  & - & 
  \frac{1}{\sqrt{2}} \left(1 + \frac{\cos\beta}{x_1}\right)
  \sin\beta \,\,\cos \theta_i \,\,
  \cos\left[\left(\omega^{(0)} + \omega_1^{(1)}+C\,\Omega \cos \beta\right) t + \phi\right]
  \nonumber \\
  & - & 
  \frac{1}{\sqrt{2}} \left(1 + \frac{1+\cos\beta}{x_1} \right)
  \sin^2\frac{\beta}{2} \,\,\sin \theta_i \,\,
  \cos\left[\left(\omega^{(0)} + \omega_1^{(1)}+C\,\Omega \cos \beta + \Omega\right) t + \phi\right]
  \label{ampl}
\end{eqnarray}

\normalsize
More generally for arbitrary $\ell$ and $m$, we define $x_{|m|}$ 
\begin{equation}
  x_{|m|} \equiv \frac{\omega^{(1)}_0 - \omega^{(1)}_{|m|}}{C \,\,\Omega},
\end{equation}
where $C$ is the rotational splitting coefficient for the appropriate
$\ell$ values.  With these definitions, the luminosity perturbations
associated with $\ell=2$ oblique pulsation are given below:
\\


$\ell=2$, $m=-2$:
\begin{eqnarray}
  \Delta L/L  & = &
  \sqrt{\frac{3}{8}} \sin^4\frac{\beta}{2} \,\, \sin^2 \theta_i 
  \left(1 + \frac{2 (1+\cos \beta)}{x_1-x_2} \right)
  \cos\left[\left(\omega^{(0)} + \omega_2^{(1)}- 2 C\,\Omega \cos \beta - 2 \Omega\right) t + \phi\right]
  \nonumber \nonumber \\
  & + & 
  \sqrt{\frac{3}{8}} \sin^2\frac{\beta}{2} \,\, \sin \beta \,\, \sin 2\theta_i 
  \left(1 + \frac{1+2 \cos \beta}{x_1-x_2} \right)
  \cos\left[\left(\omega^{(0)} + \omega_2^{(1)}- 2 C\,\Omega \cos \beta - \Omega\right) t + \phi\right]\nonumber  \\
  & + & 
  \frac{1}{8}\sqrt{\frac{3}{2}} \sin^2 \beta \,\, \left(1+3\cos 2\theta_i\right)
  \left(1 + \frac{2 \cos \beta}{x_1-x_2} \right)
  \cos\left[\left(\omega^{(0)} + \omega_2^{(1)}- 2 C\,\Omega \cos \beta \right) t + \phi\right] \nonumber \\
  & - & 
  \sqrt{\frac{3}{8}} \cos^2\frac{\beta}{2} \,\,\sin \beta \,\,\sin 2\theta_i 
  \left(1 + \frac{-1+2 \cos \beta}{x_1-x_2} \right)
  \cos\left[\left(\omega^{(0)} + \omega_2^{(1)}- 2 C\,\Omega \cos \beta + \Omega\right) t + \phi\right] \nonumber \\
  & + & 
  \sqrt{\frac{3}{8}} \cos^4\frac{\beta}{2} \,\, \sin^2 \theta_i 
  \left(1 + \frac{2 (-1+\cos \beta)}{x_1-x_2} \right)
  \cos\left[\left(\omega^{(0)} + \omega_2^{(1)}- 2 C\,\Omega \cos \beta + 2 \Omega\right) t + \phi\right] 
\end{eqnarray}

$\ell=2$, $m=-1$:
\begin{eqnarray}
  \Delta L/L  & = &
  \frac{1}{8}\sqrt{\frac{3}{2}} \sin\beta\,\, \sin^2\theta_i 
  \left(2-2\cos\beta - \frac{3\sin^2\beta}{x_1} - \frac{(1-\cos\beta)^2}{x_1-x_2} \right)
  \cos\left[\left(\omega^{(0)} + \omega_1^{(1)}- C\,\Omega \cos \beta - 2 \Omega\right) t + \phi\right]
  \nonumber \\
  & + & 
  \frac{1}{4}\sqrt{\frac{3}{2}} \sin 2\theta_i 
  \left[1+\cos\beta-2\cos^2\beta + \sin^2 \beta \left(-\frac{3\cos\beta}{x_1} + \frac{-1+\cos\beta}{x_1-x_2} \right) \right]
  \cos\left[\left(\omega^{(0)} + \omega_1^{(1)}- C\,\Omega \cos \beta - \Omega\right) t + \phi\right] 
  \nonumber \\
  & + & 
  \frac{1}{16}\sqrt{\frac{3}{2}} \sin \beta \,\, \left(1+3\cos 2\theta_i\right)
  \left[4\cos\beta-\frac{1+3\cos 2\beta}{x_1} - \frac{2 \sin^2\beta}{x_1-x_2} \right]
  \cos\left[\left(\omega^{(0)} + \omega_1^{(1)}- C\,\Omega \cos \beta \right) t + \phi\right] 
  \nonumber \\
  & - & 
  \frac{1}{4}\sqrt{\frac{3}{2}} \sin 2\theta_i 
  \left[-1+\cos\beta+2\cos^2\beta + \sin^2 \beta \left(\frac{3\cos\beta}{x_1} - \frac{1+\cos\beta}{x_1-x_2} \right) \right]
  \cos\left[\left(\omega^{(0)} + \omega_1^{(1)}- C\,\Omega \cos \beta + \Omega\right) t + \phi\right] 
  \nonumber \\
  & - & 
  \frac{1}{8}\sqrt{\frac{3}{2}} \sin\beta\,\, \sin^2\theta_i 
  \left(2+2\cos\beta + \frac{3\sin^2\beta}{x_1} + \frac{(1-\cos\beta)^2}{x_1-x_2} \right)
  \cos\left[\left(\omega^{(0)} + \omega_1^{(1)}- C\,\Omega \cos \beta + 2 \Omega\right) t + \phi\right]
\end{eqnarray}

$\ell=2$, $m=0$:
\begin{eqnarray}
  \Delta L/L  & = &
  \frac{3}{8} \sin^2\beta\,\, \sin^2\theta_i 
  \left( 1 + \frac{2}{x_1}\right)
  \cos\left[\left(\omega^{(0)} + \omega_0^{(1)} - 2 \Omega\right) t + \phi\right]
  \nonumber \\
  & + &
  \frac{3}{8} \sin 2\beta\,\, \sin 2\theta_i 
  \left( 1 + \frac{1}{x_1}\right)
  \cos\left[\left(\omega^{(0)} + \omega_0^{(1)} - \Omega\right) t + \phi\right]
  \nonumber \\
  & + &
  \frac{1}{16} \left(1+3\cos 2\beta\right) \left(1+3 \cos 2\theta_i\right) 
  \cos\left[\left(\omega^{(0)} + \omega_0^{(1)}\right) t + \phi\right]
  \nonumber \\
  & + &
  \frac{3}{8} \sin 2\beta\,\, \sin 2\theta_i 
  \left( 1 - \frac{1}{x_1}\right)
  \cos\left[\left(\omega^{(0)} + \omega_0^{(1)} + \Omega\right) t + \phi\right]
  \nonumber \\
  & + &
  \frac{3}{8} \sin^2\beta\,\, \sin^2\theta_i 
  \left( 1 - \frac{2}{x_1}\right)
  \cos\left[\left(\omega^{(0)} + \omega_0^{(1)} + 2 \Omega\right) t + \phi\right]
\end{eqnarray}

$\ell=2$, $m=+1$:
\begin{eqnarray}
  \Delta L/L  & = &
  \frac{1}{8}\sqrt{\frac{3}{2}} \sin\beta\,\, \sin^2\theta_i 
  \left(2+2\cos\beta - \frac{3\sin^2\beta}{x_1} - \frac{(1+\cos\beta)^2}{x_1-x_2} \right)
  \cos\left[\left(\omega^{(0)} + \omega_1^{(1)}- C\,\Omega \cos \beta - 2 \Omega\right) t + \phi\right]
  \nonumber \\
  & + & 
  \frac{1}{4}\sqrt{\frac{3}{2}} \sin 2\theta_i 
  \left[-1+\cos\beta+2\cos^2\beta + \sin^2 \beta \left(-\frac{3\cos\beta}{x_1} + \frac{1+\cos\beta}{x_1-x_2} \right) \right]
  \cos\left[\left(\omega^{(0)} + \omega_1^{(1)}- C\,\Omega \cos \beta - \Omega\right) t + \phi\right] 
  \nonumber \\
  & - & 
  \frac{1}{16}\sqrt{\frac{3}{2}} \sin \beta \,\, \left(1+3\cos 2\theta_i\right)
  \left[4\cos\beta+\frac{1+3\cos 2\beta}{x_1} + \frac{2 \sin^2\beta}{x_1-x_2} \right]
  \cos\left[\left(\omega^{(0)} + \omega_1^{(1)}- C\,\Omega \cos \beta \right) t + \phi\right] 
  \nonumber \\
  & - & 
  \frac{1}{4}\sqrt{\frac{3}{2}} \sin 2\theta_i 
  \left[1+\cos\beta-2\cos^2\beta + \sin^2 \beta \left(\frac{3\cos\beta}{x_1} + \frac{1-\cos\beta}{x_1-x_2} \right) \right]
  \cos\left[\left(\omega^{(0)} + \omega_1^{(1)}- C\,\Omega \cos \beta + \Omega\right) t + \phi\right] 
  \nonumber \\
  & + & 
  \frac{1}{8}\sqrt{\frac{3}{2}} \sin\beta\,\, \sin^2\theta_i 
  \left(-2+2\cos\beta - \frac{3\sin^2\beta}{x_1} - \frac{(1-\cos\beta)^2}{x_1-x_2} \right)
  \cos\left[\left(\omega^{(0)} + \omega_1^{(1)}- C\,\Omega \cos \beta + 2 \Omega\right) t + \phi\right]
\end{eqnarray}

$\ell=2$, $m=+2$:
\begin{eqnarray}
  \Delta L/L  & = &
  \sqrt{\frac{3}{8}} \cos^4\frac{\beta}{2} \,\, \sin^2 \theta_i 
  \left(1 + \frac{2 (1-\cos \beta)}{x_1-x_2} \right)
  \cos\left[\left(\omega^{(0)} + \omega_2^{(1)}- 2 C\,\Omega \cos \beta - 2 \Omega\right) t + \phi\right]
  \nonumber \nonumber \\
  & - & 
  \sqrt{\frac{3}{8}} \cos^2\frac{\beta}{2} \,\, \sin \beta \,\, \sin 2\theta_i 
  \left(1 + \frac{1-2 \cos \beta}{x_1-x_2} \right)
  \cos\left[\left(\omega^{(0)} + \omega_2^{(1)}- 2 C\,\Omega \cos \beta - \Omega\right) t + \phi\right]\nonumber  \\
  & + & 
  \frac{1}{8}\sqrt{\frac{3}{2}} \sin^2 \beta \,\, \left(1+3\cos 2\theta_i\right)
  \left(1 - \frac{2 \cos \beta}{x_1-x_2} \right)
  \cos\left[\left(\omega^{(0)} + \omega_2^{(1)}- 2 C\,\Omega \cos \beta \right) t + \phi\right] \nonumber \\
  & + & 
  \sqrt{\frac{3}{8}} \sin^2\frac{\beta}{2} \,\,\sin \beta \,\,\sin 2\theta_i 
  \left(1 - \frac{1+2 \cos \beta}{x_1-x_2} \right)
  \cos\left[\left(\omega^{(0)} + \omega_2^{(1)}- 2 C\,\Omega \cos \beta + \Omega\right) t + \phi\right] \nonumber \\
  & + & 
  \sqrt{\frac{3}{8}} \sin^4\frac{\beta}{2} \,\, \sin^2 \theta_i 
  \left(1 - \frac{2 (1+\cos \beta)}{x_1-x_2} \right)
  \cos\left[\left(\omega^{(0)} + \omega_2^{(1)}- 2 C\,\Omega \cos \beta + 2 \Omega\right) t + \phi\right] 
\end{eqnarray}

\end{document}